\documentclass[numberedappendix]{emulateapj}

\def\puastro{1}
\def\umcp{2}

\usepackage{amssymb} 
\usepackage{graphicx}
\usepackage{verbatim}

\newcommand{\be}{\begin{equation}}
\newcommand{\ee}{\end{equation}}
\newcommand{\ud}{\mathrm{d}}

\begin{document}  

\title{Growth of structure seeded by primordial black holes}

\author{
Katherine J. Mack\altaffilmark{\puastro},
Jeremiah P. Ostriker\altaffilmark{\puastro},
and Massimo Ricotti\altaffilmark{\umcp}
}

\altaffiltext{\puastro}{Princeton University Department of Astrophysical
Sciences, Peyton Hall - Ivy Lane, Princeton, NJ 08544}

\email{mack@astro.princeton.edu}

\altaffiltext{\umcp}{Department of Astronomy, University of Maryland
College Park, MD 20742}


\begin{abstract}

We discuss the possibilities for primordial black holes
(PBHs) to grow via the accretion of dark matter.  In
agreement with previous works, we find that accretion
during the radiation-dominated era does not lead to a
significant mass increase.  However, during
matter-domination, PBHs may grow by up to two orders of
magnitude in mass through the acquisition of large
dark matter halos.  We discuss the possibility of PBHs
being an important component in dark matter halos of
galaxies as well as their potential to explain the
ultra-luminous x-ray sources (ULXs) observed in nearby
galactic disks.  We point out that although PBHs are
ruled out as the dominant component of dark matter,
there is still a great deal of parameter space open to
them playing a role in the modern-day universe.  For
example, a primordial halo population of PBHs each at
$10^{2.5} M_\odot$ making up $0.1\%$ of the
dark matter grow to $10^{4.5} M_\odot$
via the accumulation of dark matter halos to account for
$\sim 10\%$ of the dark matter mass by a redshift of
$z \approx 30$.  These intermediate mass black holes
may then ``light up'' when passing through molecular
clouds, becoming visible as ULXs at the present day,
or they may form the seeds for supermassive black
holes at the centers of galaxies.

\end{abstract}



\section{Introduction}
\label{sec:intro}

It is well-established that supermassive black holes
(SMBHs) with masses in the range $m_{BH}
\sim 10^6 - 10^{9.5} M_{\odot}$ reside in the centers of
spheroidal systems \citep{Bernardi2003}.  One can make a
convincing case that these have grown largely through
accretion, with the consequent energy emission observed
in electromagnetic output and jets at an efficiency
of $\epsilon \sim 0.1$ \citep{Soltan,YuTremaine}.
Observations of distant quasars have shown us that these
SMBHs are already in place by redshifts of 6 and
greater, but the mechanism of their formation remains a
mystery.

Motivated by this question, we examine the potential for
primordial black holes (PBHs) to grow through accretion
to become seed masses for SMBHs.  Primordial black holes,
defined as black holes forming in the early universe without
stellar progenitors, were first proposed by \citet{ZN1967}
and \citet{Hawking1971} as a possible consequence of the
extremely high densities achieved in the Big Bang model.
If they do indeed form in the early universe and can avoid
evaporation~\citep{Hawking1975} up to the present day, they
must still exist and they may be important.  We use a
combination of analytical and numerical methods to follow
PBH growth through the radiation- and matter-dominated
eras and show that a PBH can multiply its mass by up to
two orders of magnitude through the accretion of a dark
matter halo.

If PBHs can grow sufficiently by accretion (or if they are
very large at birth)
they may account for intermediate-mass black holes
(IMBHs) in the mass range
$10^2 M_\odot \lesssim M \lesssim 10^4 M_\odot$, which have been
suggested as the engines behind ultra-luminous x-ray
sources (ULXs) recently discovered in nearby galactic disks
\citep{Dewangan2005,Madhusudhanetal2006,Miller2004,Mushotzky2004}.
Based on the observed ULX luminosities of $\sim 10^{39} 
\textrm{erg s}^{-1}$, stellar mass black holes are ruled
out unless the emission is highly beamed.  IMBHs have
the appropriate
Eddington luminosity to explain ULXs, but since there is
currently no easy way to produce black holes of this
mass from stellar collapse at the abundances observed
\citep{FryerKalogera2001},
their origins are highly
debated.  We suggest that PBHs may be able to grow to
sufficient masses through the accretion of dark matter
halos to account for a pervasive population of IMBHs.

These IMBHs may also be important in the build-up of SMBHs,
as suggested by recent numerical
studies \citep{MicicAbelSigurdsson2005}, a scenario that
may be testable with the Laser Interferometer Space Antenna
(LISA) in coming years \citep{Micic2,Clifford2004,Fregeau2006}.

Aside from observational indications,
there have been physics-based inquiries indicating the
plausibility of the production of
PBHs through a variety of mechanisms in the early
universe \citep{Carr2005}, as briefly reviewed in \S
\ref{sec:PBH}.

One may wish to ask at this point what constraints on PBH
production exist given current observational limits.  This
important question is addressed in \S~\ref{sec:PBHobslims}
and Figure~\ref{constraints}.  We will show that there is
ample room for a population of PBHs that is both permitted
and interesting.

Our study of PBH accretion is organized as follows.
In \S \ref{sec:PBH} we discuss theories of
PBH formation.  In
\S \ref{sec:acc} we outline our accretion model.
Sections \ref{sec:rad} and \ref{sec:matt} describe the
accretion calculations in the radiation and matter eras,
respectively, while in \S \ref{sec:comb} we do a combined
calculation for both eras.  In \S \ref{sec:results} we
give our results for the total accretion possible for a
PBH, and in \S \ref{sec:discussion} we discuss the
implications of our findings for the possible importance
of PBHs in the present-day universe.  Appendices
\ref{sec:apprad} and \ref{sec:appmatt} present details
of our accretion calculations.

\section{Primordial Black Hole Production}
\label{sec:PBH}

There has been a great deal of interest in primordial
black hole production in the early universe, resulting
in the proposal of a variety of formation mechanisms.
We refer the reader to a review of PBHs \citep{Carr2005}
for an overview of the possibilities, briefly summarized
here.  In one mechanism of interest
\citep{Jedamzik1997}, PBHs form at a QCD phase
transition at $\sim 1 M_\odot$, a scale
of interest for microlensing studies.  PBHs formed
at higher mass through other mechanisms may be natural
candidates to solve other problems, such as that of the
nature of ULX engines.  In general, most mechanisms create
PBHs at about the horizon mass, given by \citep{Carr2005}:
\be
M_H(t) \approx \frac{c^3t}{G},
\ee
or in terms of cosmic temperature T,
\be
M_H(T) \approx 1 M_\odot \left(
\frac{T}{100 ~\textrm{MeV}} \right)^{-2} \left(
\frac{g_{\textrm{eff}}}{10.75} \right)^{-1/2},
\ee
where $g_{\textrm{eff}}$ is the number of effective
relativistic degrees of freedom.

Briefly, some common mechanisms for creation of PBHs
in the early universe are: (1) PBHs formed at the QCD
phase transition (mentioned above), when conditions temporarily allow
regions of modest overdensity to collapse into black
holes when they enter the horizon \citep{Jedamzik1997}.
The PBHs formed in this way would have a mass spectrum
strongly peaked at the QCD epoch horizon mass
($\sim 1 M_\odot$). (2) The collapse of rare peaks in
the density field of the early universe.  In this case,
the probability of PBH formation at a given epoch is
determined by the nature and evolution of the
perturbations.  (3) The collapse of
cosmic string loops \citep{CaldCasp1996,GarSak1993,
Hawking1989,PolnarevZemboricz1988,MacBrandWich1998}.
Due to frequent collisions and
reconnections, cosmic strings may occasionally form
loops compact enough that the loop is within its
Schwarzschild radius in every dimension.  (4) A soft
equation of state \citep{KhlopovPolnarev1980}.
If the equation of state becomes soft
(e.g., during a phase transition), PBHs may form at peaks
in density as pressure support weakens.  (5) Bubble
collisions \citep{CrawfordSchramm1982,HawkingMS1982,
LaStein1989}.  During spontaneous symmetry breaking, bubbles
of broken symmetry may collide, in some cases focusing
energy at a point and producing a black hole.  In this
mechanism, the PBHs would form at the horizon mass of the
phase transition.  (6) Collapse of domain walls
\citep{Berezin1983,IpserSikivie1984}.  Closed
domain walls forming at a second-order phase transition
may collapse to form PBHs.  In the case of thermal
equilibrium, this would result in very small masses, but
see \citep{Rubin2000} for a discussion of how
non-equilibrium conditions may result in significant PBH
masses.

For cases in which the mass of the PBH is low at creation,
the PBH may evaporate before the present day through Hawking
radiation \citep{Hawking1975}.  The limiting mass for
evaporation by the present day is $10^{15}$ g; in general
the evaporation time is given by \citep{Carr2003}
\be
\tau(M) \approx \frac{\hbar c^4}{G^2 M^3} \approx 10^{64}
\left( \frac{M}{M_\odot} \right)^3 \textrm{yr}.
\ee

If PBHs are to arise directly from primordial density
perturbations, it is required that the scale of fluctuations
set down by inflation be ``blue'' -- i.e., the spectrum must
have more power on small scales.  In terms of inflationary
parameters, this implies that the scalar spectral index
$n>1$, which is disfavored in the latest WMAP results
\citep{Spergel2006}.

Some mechanisms, such as the collapse of density peaks, may
result in PBHs forming in clusters.  For a discussion of the
consequences of clustering, see \citep{Chisholm2005}.
Formation via domain wall collapse, as discussed in
\citet{Dokuchaev2004}, may also lead to clustering, without
relying on initial dark matter perturbations.  In that scenario,
primordial black holes
can grow through mergers to form galaxies without the help of
initial perturbations in the dark matter.

In this work, we assume the PBHs are rare and isolated rather
than appearing in clusters, but we expect that clustered PBHs
would increase accretion power, so in that respect our
treatment is a conservative one.  We may refer to a specific
PBH seed mass when convenient for illustrative purposes, but
it should be noted that our results are independent of the
PBH formation mechanism.

\section{Accretion model}
\label{sec:acc}
\subsection{Setup}

We model accretion of matter onto primordial black holes in
both the radiation era and the matter era.  In both cases we
follow the calculations for radial infall, following
previous work on the growth of clusters \citep{GG,Bert,
FillmoreGoldreich}.
Acknowledging that in a realistic accretion model the
infall is unlikely to be perfectly radial, we make the 
simplifying assumption that the angular momentum of the
infalling matter causes it to accrete in a halo around the
PBH rather than incorporating itself into the PBH itself.
This assumption is conservative from the standpoint of an
estimation of the PBH's mass increase.

A PBH clothed by a
dark matter halo will have the accreting power of an object
having the total mass of the PBH plus the halo, to the
extent that the accretion radius (e.g., Bondi radius) is
larger than the radius of the PBH dark matter halo.
However, constraints on the PBH's effect on the power
spectrum apply only to the seed mass, not to the total mass
of the clothed PBH, since the mass accreted by the PBH
is drawn from the surrounding matter, and the additional
mass is therefore ``compensated.''  In other words, a
region may be defined around the PBH for which the
overdensity is due only to the original PBH, with no
contribution from the accreted mass.

\subsection{Assumptions}

In all cases, we use the cosmological parameters derived
from the third-year WMAP data release \citep{Spergel2006}.
Specifically, we use the parameter set derived from the
assumption of a flat, $\Lambda$CDM universe, with the
combination of WMAP III and all other data sets
($\Omega_\Lambda = 0.738, \Omega_m = 0.262, h = 0.708, 
\sigma_8 = 0.751$).  In both
the radiation and matter era calculations, we consider the
accretion of dark matter only.  In the radiation era, we
assume the radiation is too stiff to accrete
appreciably, as suggested in many past analyses
\citep{CarrHawking,CustodioHorvath,NiemeyerJedamzik}.  In
the matter era, we ignore the small contribution to the PBH
mass due to the accretion of baryons.

We make the further assumptions that each PBH is stationary
and isolated, and that the surrounding matter is initially
in the Hubble flow.

In all our accretion models, we end the calculation
at a sufficiently
high redshift that the effect of the cosmological
constant is negligible.  For completeness, however, we
include in Appendix \ref{sec:appmatt} the outline of
the calculation with the cosmological constant
included.

For a more detailed analysis of the consequences of
gas accretion onto PBHs, we refer the reader to a
companion paper \citep{ROM}.

\subsection{PBH velocities}

\begin{figure}[htb]
\begin{center}
\resizebox{\columnwidth}{!}{\includegraphics[angle=90]{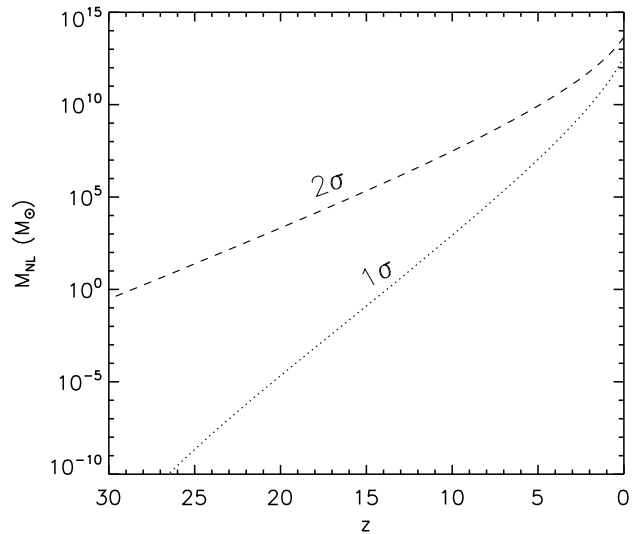}}
\end{center}
\caption{Mass of 1-$\sigma$ fluctuations (dotted line) and
2-$\sigma$ fluctuations (dashed line) as a function of
redshift.}\label{NLmass}
\end{figure}

Our accretion estimate would decrease if the PBHs were
moving quickly relative to the dark matter surrounding
them; here, we assume that the PBHs are initially stationary.
We justify this assumption by considering the likely
effect of nearby density perturbations in the dark matter.
At any epoch, we can estimate the mass scale at which
structures are becoming nonlinear by calculating the
variance of density perturbations from an estimate of
the matter power spectrum.  In Figure~\ref{NLmass}, we
plot the nonlinear mass scales for $1$ and $2\sigma$
perturbations.  The fitting formula
\be
M_{2\sigma} = (1 \times 10^{17} M_\odot) \exp(-5.57(1+z)^{0.57})
\ee
approximates the $2\sigma$ mass perturbations.  From
this, we may calculate the characteristic circular velocity
and thus the typical proper velocity of PBHs as a function of
redshift:
\be
v_p \sim v_c = (17~\textrm{km s}^{-1}) \left(\frac{M_{2\sigma}}
{10^8 M_\odot}\right)^{1/3} \left(\frac{1+z}{10}\right)^{1/2}.
\ee
For redshifts down to $z \sim 30$, the peculiar velocities
are low and we can consider the PBHs to be stationary.

\section{Radiation era}
\label{sec:rad}

The details of an analytical estimate of accretion in the
radiation era can be found in Appendix \ref{sec:apprad}.
Here we outline the basic idea and quote the result of a
numerical calculation.

We begin the calculation at a redshift $z \approx 10^7$.
In the radiation era, the motion of a dark matter shell
a distance $r$ from the black hole is governed by the
differential equation
\be
\frac{\ud^2 r}{\ud t^2} = -\frac{G m_{BH}}{r^2}-\frac{1}{4}\frac{r}{t^2},
\ee
where $m_{BH}$ is the black hole mass and $t$ is time.
With the initial conditions
\be
r = r_i, \quad \frac{\ud r}{\ud t} = H_i r_i = \frac{1}{2}\frac{r_i}{t_i}
\ee
at $t=t_i$, we evolve these equations forward in time
until matter-radiation equality at
$z_{eq} \approx 3 \times 10^3$.
When a shell turns around ($\dot{r}=0$), we assign the
matter in that shell to the PBH's dark matter halo.

We find that the PBH can accrete a dark matter halo on
order its original mass:
\be
\frac{m_{h,rad}}{m_{BH}} \approx 1.
\ee

\section{Matter era}
\label{sec:matt}

The evolution of a spherically symmetric overdensity
in the matter-dominated era has been treated in
the case of the growth of clusters
\citep{GG,Bert,FillmoreGoldreich}.
These analyses neglect the effect of the
cosmological constant, assuming $\Omega_\Lambda=0$ and
$\Omega_m = 1$.  In the general case where
$\Lambda \neq 0$ the equation of
motion of a shell of dark matter a radius $r$ from the
PBH becomes \citep{Lahav}
\be
\frac{\ud^2 r}{\ud t^2} = -\frac{G m_{BH}}{r^2} + \frac{\Lambda r}{3}.
\ee
The cosmological constant term affects the accretion at
redshifts on order 1, but since we halt our accretions at
higher redshift, we find growth consistent with the \citet{GG}
and \citet{Bert} result:
\be
m_{h} \sim t^{2/3},
\ee
with the turnaround radius of the dark matter halo
(which we will identify as the effective radius of the
halo) growing with time as
\be
r_{ta} \sim t^{8/9}.
\ee

The details of the $\Lambda \neq 0$ calculation are
discussed in Appendix \ref{sec:appmatt}.

For a PBH that begins growing at
matter-radiation equality and stops at $z=30$, we
find that the halo increases its total mass as
\be
\frac{m_{h,matter}}{m_{BH}} \approx 100.
\ee
In the general case of a PBH growing in the matter era,
the mass increase from $z_{eq}$ to $z_f$ is
\be
\frac{m_{h,matter}}{m_{BH}} \approx 100\left(\frac{31}{1+z_f}\right).
\ee

\section{General case}
\label{sec:comb}

In addition to approximate calculations specific to
the matter and radiation eras, we also present the
general result, which spans both eras and
includes (for completeness) consideration of the
cosmological constant.

We start with the radial infall equation for
a shell of matter:
\be
\frac{\ud^2 r}{\ud t^2} = \frac{-4 \pi G r}{3}
(\rho_m + 2\rho_r) + \frac{\Lambda c^2 r}{3}.
\ee
For computational convenience, we recast this
equation in terms of derivatives with respect to
redshift, and we switch to comoving coordinates.
After some algebra, we are left with two differential
equations, one for the comoving radial coordinate
$x(z)$ and one for the peculiar velocity of a shell
$v(x,z)$ defined by
\begin{eqnarray}
v & = & \frac{\ud r}{\ud t} - H r \\
  & = & \frac{\ud (ax)}{\ud t} - H a x,
\end{eqnarray}
where $a = 1/(1+z)$.
The integration equations take on a simple form in
the new coordinates:
\begin{eqnarray}
\frac{\ud x}{\ud z} & = & \frac{-v}{H} \\
\frac{\ud v}{\ud z} & = & av + \frac{G (M_{acc}(x) + m_{BH})}
{H a x^2},
\end{eqnarray}
where $M_{acc}(x)$ is defined as
the excess matter over the background-density matter
within the comoving radius $x$ (i.e., the matter
previously accreted into the halo region around the
black hole).

\section{Results}
\label{sec:results}

\subsection{Mass Accretion}

\begin{figure}[htb]
\begin{center}
\resizebox{\columnwidth}{!}{\includegraphics{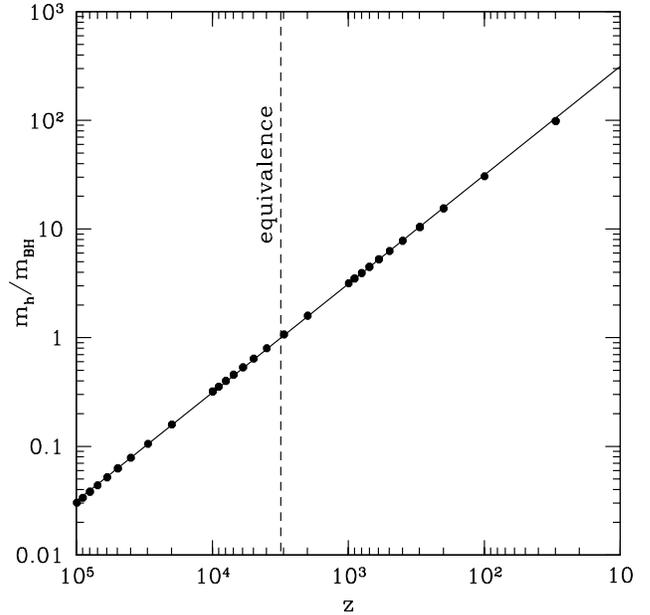}}
\end{center}
\caption{Accreted halo mass vs. redshift.  The halo radius is 
defined at an overdensity $\delta = 2$.We include a line
to indicate the redshift of matter-radiation equality.}\label{fig:acc}
\end{figure}

Our results from the combined calculation are
consistent with those we obtained treating
the matter and radiation eras separately.  A PBH
can grow by two orders of magnitude through the
accumulation of a dark matter halo from early in the
radiation era to $z \sim 30$, with the halo
mass increasing proportional to the cosmic scale
parameter $a = 1/(1+z)$:
\be
m_h(z) = \phi_i \left(\frac{1000}{1+z}\right) m_{BH},
\ee
where the proportionality constant $\phi_i \approx 3$.
Figure~\ref{fig:acc} summarizes our mass accretion result.

\subsection{Halo Profile}

\begin{figure}[htb]
\begin{center}
\resizebox{\columnwidth}{!}{\includegraphics{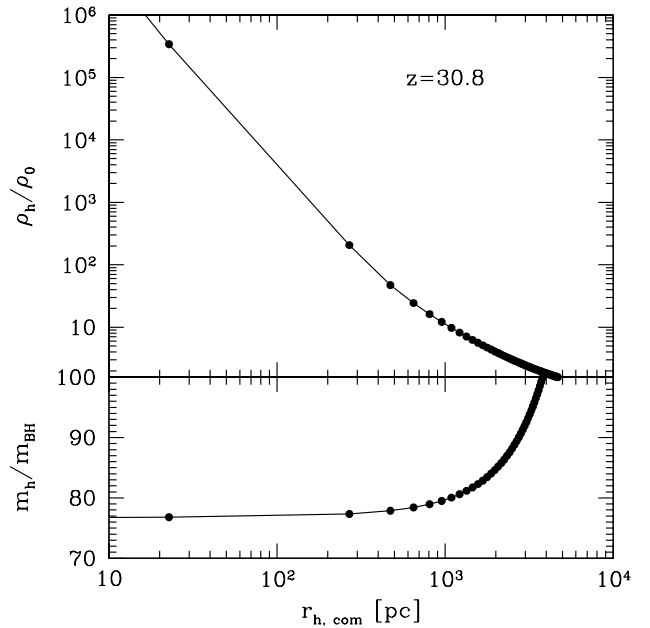}}
\end{center}
\caption{Dark matter halo profile.  Top panel: halo overdensity vs.
comoving radius from PBH; bottom panel: halo mass vs. comoving radius
from PBH.  In the inner parts of the halo, the density falls off
as $r^{-3}$, and the profile flattens in the outer 
regions. }\label{fig:profile}
\end{figure}

In a previous study, \citet{Bert}
performed analytical calculations of radial infall
onto a central overdensity and onto a black hole;
the difference in the two calculations was that in the
former case, the particles could oscillate through the
center, whereas in the latter case they were absorbed
by the black hole (as is the case in our simulation).
Bertschinger obtained a 
$\rho(r) \sim r^{-2.25}$ profile for the extended
overdensity and a $\rho(r) \sim r^{-1.5}$ profile for
the black hole case.  Our simulation resulted in a
profile of $\rho(r) \sim r^{-3}$, differing from either
of the above cases.  Our profile is illustrated in
Figure~\ref{fig:profile}.

Since a powerlaw profile has no sharp cut-off in radius, we
must choose a criterion by which to define the matter within the
halo.  We may choose either the turnaround radius (the distance
out to which shells have broken free of the Hubble flow) or a
cut on the overdensity vs. radius; both criteria give similar
results.  Using the turnaround radius definition, the comoving
radius of the dark matter halo at redshift $z_f$ from accretion
beginning at matter-radiation equality is
\be
x_{ta} = 1.30 \left(\frac{1+z_f}{31}\right)^{-1/3} 
\left(\frac{m_{BH}}{100 M_\odot}\right) \rm{kpc}
\ee
for $z_f \lesssim 100$.

The radius defined by a cut on overdensity can be read off
Figure~\ref{fig:profile}.

\subsection{Density Parameter in PBHs}

As the masses of the clothed PBHs increase, so does
their overall density parameter.  Given an initial
matter fraction 
\be
\omega_{BH,i} \equiv \frac{\Omega_{BH,i}}{\Omega_{m,i}},
\ee
the final matter fraction increases in proportion to
the clothed PBH mass:
\be
\frac{\omega_{BH,f}}{\omega_{BH,i}} = \frac{m_{BH,f}}{m_{BH,i}}
\ee
where $m_{BH,f}$ includes the PBH and the accreted halo.

\begin{figure}[htb]
\begin{center}
\resizebox{\columnwidth}{!}{\includegraphics[angle=90]{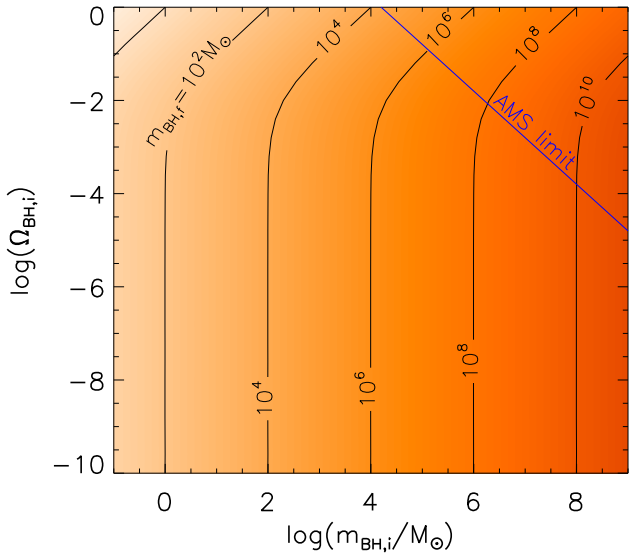}}
\end{center}
\caption{Final mass per PBH plotted as a function of
initial mass and initial PBH density parameter, with
the \citet{AMS2003} limit included for
reference (see \S~\ref{sec:PBHobslims}).}\label{contour}
\end{figure}

In Figure~\ref{contour}, we illustrate that the
proportional mass increase, while not dependent on the
mass of the PBH, does depend on the proportion of the
dark matter made up of PBHs.  When the PBHs begin to
dominate the dark matter, they grow less because of the
decrease in the density of dark matter.

\section{Discussion \& Conclusions}
\label{sec:discussion}

Our results suggest that PBHs could grow significantly after
their formation by acquiring a dark matter halo and the
resulting clothed black holes could
make up an interesting fraction of the dark matter.  In the
following discussion, we show that current observations are
not in conflict with this conclusion, and in fact there is
ample room both observationally and theoretically for PBHs
to play a role in the universe today.

\subsection{Observational Limits on PBHs}
\label{sec:PBHobslims}

\begin{figure}[htb]
\begin{center}
\resizebox{\columnwidth}{!}{\includegraphics{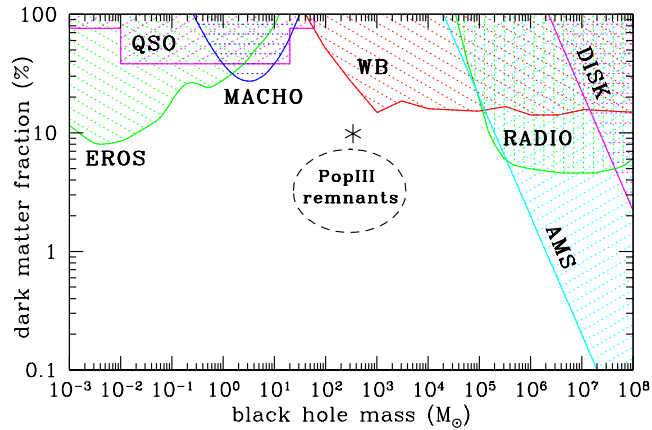}}
\end{center}
\caption{\label{constraints}Observational constraints on black
holes in the Galactic halo (see \S \ref{sec:PBHobslims}) from
microlensing experiments (EROS
and MACHO), quasar variability studies (QSO), compact radio
source lensing (RADIO), the stability of wide binaries (WB),
the high-wavenumber matter power spectrum (AMS), and the heating
of the Galactic disk (DISK).  The region labeled ``PopIII remnants''
represents a rough estimate of the region of parameter space
 relevant to a
scenario in which Population III star remnant IMBHs are
responsible for ULX observations.  The star symbol gives the
position of the example of a PBH ULX used in the abstract (see
\S \ref{sec:ULX}).}
\end{figure}

We include in Figure~\ref{constraints} a plot of the current
observational limits on PBHs over a wide range of masses and
dark matter fractions.  Here, we describe the limits
illustrated in the plot.

Most limits on PBHs in the
present-day universe are derived from considerations of PBHs
as dark matter candidates.  PBHs massive enough to escape
evaporation would certainly qualify as ``cold'' and ``dark''
matter; however, their existence could have noticeable effects
on processes from
nucleosynthesis to galaxy formation.  If PBHs form at
very early times, and thus with very low masses
($\lesssim10^{10}$ g),
they could interfere with nucleosynthesis by emitting
particles during their evaporation \citep{Kohri2000}.  The
abundance of PBHs forming after nucleosynthesis is
constrained by measurements of the baryon fraction of the
universe \citep{Novikov1979}.  Massive PBHs can also be
constrained by
dynamical considerations in the low-redshift universe.  Some
of the strongest current constraints are derived from the wealth
of data from microlensing searches in Galaxy \citep{Alcock2000,
Alcock2001, Afonso2003}.  The frequency and character of observed
microlensing events constrains black holes in the mass range
0.1 - 1 $M_\odot$ to make up less than $\sim 20$\% of the dark matter
in the Galactic halo \citep{Alcock2000,Gould2005}.  The limits
from microlensing are shown on Figure~\ref{constraints} labeled
``MACHO'' and ``EROS.''

For larger masses, constraints on PBH dark matter can be found
by examining the effect of PBHs on the matter power spectrum.
Calculating the excess in the power spectrum that
Poisson-distributed PBHs would contribute, 
\citet{AMS2003} find an upper 
limit on present-day PBH mass of a few times $10^4 M_\odot$,
which is consistent with but tighter than the previous constraint
at $10^6 M_\odot$ based on the heating
of the Galactic disk \citep{LaceyOstriker}.  However, we
note that this limit assumes all the dark matter is in PBHs:
$\Omega_{BH} = \Omega_{DM}$.  If this assumption is relaxed,
we find that a wide range of masses can be accommodated at lower
density parameters.  Specifically, we find that the product of
the initial density parameter and the initial mass are
constrained by
\be \label{eqn:AMS}
m_{BH} \Omega_{BH} < \frac{m_{AMS}}{x}
\ee
where $x$ is the factor by which the PBHs increase in mass via
accretion and $m_{AMS}$ is $\sim 10^4 M_\odot$.  We include this
limit in the constraint plot (Figure~\ref{constraints}) with the
label ``AMS.''

Other limits can be placed by considering the effects of
compact objects along the line of sight lensing more distant
sources \citep{Wambsganss2002}.  \citet{Dalcantonetal1994}
search for
the slight amplification in the continuum emission of quasars
that would be expected were black holes to cross the line of
sight during an observation.  With a large sample of
observations, they are able to place limits on black holes in
the range $\sim 10^{-3} M_\odot$ to $\sim 300 M_\odot$.  This
limit is included in the constraints plot
(Figure~\ref{constraints}) and labeled ``QSO.''
\citet{Wilkinsonetal2001} place
limits on the abundance of massive black holes in the universe
based on their predicted effect of creating multiple images of
compact radio sources.  Studying a sample of 300 sources, they
find a null result and from that can place a constraint on the
density of black holes along the line of sight.  Their
constraint is included in Figure~\ref{constraints} with the
label ``RADIO.''  Coincident with the compact radio source
study, another group found a similar constraint by searching for
the same lensing effect in gamma ray burst light
curves \citep{Nemiroffetal2001}.  The results of the two studies
are consistent with each other, so for simplicity we include
only the \citet{Wilkinsonetal2001} result in
Figure~\ref{constraints}.

Finally, a limit on black holes with masses of $\sim 10 M_\odot$
and up can be placed by observing widely orbiting binary systems
in the Galaxy \citep{YooChanameGould2004}.  If a compact object
passes between the two companion stars in a binary, the orbits of
the stars will be perturbed.  \citet{YooChanameGould2004} use this
to estimate how many compact objects with the ability to disturb
a binary system exist in the halo.  This constraint is included in
Figure~\ref{constraints} and labeled ``WB.''

We point out that none of the above observations
significantly constrain black holes making up less than $\sim$
10\% of the dark matter for a wide range of masses.

\subsection{PBHs as ULXs}\label{sec:ULX}

This part of parameter space is consistent with an
interpretation of the recent ULX observations as accreting,
intermediate-mass black holes in nearby galaxies.  In this
scenario, ULXs occur when IMBHs residing in the galaxy's
halo pass through molecular clouds in the
disk \citep{Mapellietal2006,Miller2004,Winteretal2006}. 
The enhanced density in a molecular cloud is sufficient to
trigger gas accretion, which causes the IMBH to emit X-ray
radiation.  These sources would be transient, and in any
given galaxy the number of sources detected at any given
time would depend on the number and distribution of the
IMBH population and the fraction of the disk made up of
molecular clouds.

Several recent papers explore the possibilities for making
ULXs with IMBHs.  \citet{MiiTotani2005}
estimate the number of ULXs expected if they are the result
of IMBHs passing through
molecular clouds.  Using the mass and dark matter fraction
estimated for IMBHs if they are the compact remnants of
Population III stars \citep{MadauSilk2005}, $M_{IMBH} \sim
10^2 - 10^3$ and $\Omega_{IMBH}/\Omega_b \sim 0.1$, they
find that the estimated number of ULXs is consistent with
observations.  These results do not depend on the nature
of the IMBHs -- if the IMBHs were primordial in nature rather
than Pop III star remnants, they would also be capable of
producing the observed sources.

Although these results are encouraging, the present lack of
understanding of the nature of
ULXs and the many uncertainties that go into predictions
of the consequences of a halo population of IMBHs mean that
the issue is far from resolved.  A recent paper drawing on the
Mii \& Totani result uses an ensemble of N-body simulations
to place constraints on the number of IMBHs in the Milky Way
halo by drawing on the fact that we have not observed any
ULXs in our galaxy \citep{Mapellietal2006}.  These authors
find in one simulation that for a halo population of
$\sim 10^5$ IMBHs incorporating a fraction $0.1$\% of
the baryons and distributed in an NFW profile \citep{NFW},
the predicted number of ULXs per galaxy is on order 1;
however,
the number of lower-luminosity X-ray sources is overproduced.
The constraint found by this method may be applicable to our
own galaxy, but it has not as yet been extended to other
galaxies, where ULXs are observed.  Since the predictions
depend strongly on not only the distribution and number of
IMBHs, but also on the properties of the gas in the galactic
disk and the efficiency of the black hole accretion (which
in turn depends on whether or not an accretion disk is
formed), it is difficult to generalize them to other systems.

Other authors have suggested that ULXs may be due to IMBHs
accreting from captured stellar companions rather than
molecular clouds \citep{PooleyRappaport2005,Patrunoetal2006,
Madhusudhanetal2006}.  In this case, the ULXs would
``turn on'' when residing in dense star clusters.

More detailed observations and simulations are required
to answer the ULX question.  Here, we merely point
out that the case for IMBH ULXs is an interesting one,
easily consistent with current constraints on the dark
matter fraction in black holes, and as we show, PBHs may
account for or grow into IMBHs by the present era.  Thus
PBHs should be considered viable candidates to explain
these mysterious sources.

In Figure~\ref{constraints}, we include as a region
of interest the area in parameter space explored in the
Mii \& Totani \citep{MiiTotani2005} paper
(labeled ``PopIII remnants'').  We also mark with an asterisk
the position of the scenario discussed in the abstract:
a population of $10^{2.5} M_\odot$ PBHs making up $0.1$\%
of the dark matter and growing through accretion to
incorporate $10$\% of the dark matter by the present day
(we mark the mass of the seed PBH only, going on the
conservative assumption that the PBH's dark matter halo is
not directly accreted).  Both these points lie in a region
of acceptable parameter space for the explanation of ULXs
with IMBHs, but the exact extent of the region is difficult
to define given the uncertainties mentioned above.

\subsection{PBHs as SMBH Seeds}

The question of the origin of supermassive black holes at high
redshifts has attracted a great deal of attention since quasars
have been discovered at redshifts $z>6$, implying that
black holes as massive as $\sim 10^9 M_\odot$ \citep{Fan2003,
Barthetal2003,Willottetal2003} exist when the universe is less than
1 Gyr old.  It has proven difficult to find a mechanism that can
create such massive black holes so quickly.  Most proposals
require smaller black holes to act as seeds for the build-up of
SMBHs.  In some cases, these seeds form directly from the collapse
of halos \citep{LodatoNatarajan2006,SpaansSilk2006}.  In others,
the seeds are the remnants of Population III stars
\citep{Shapiro2005,VR}, or they might be
primordial. The question of whether or not SMBHs can be grown
from the merging of stellar mass black holes has also been
discussed in recent work
\citep[see][and references therein]{Lietal2006}.
While in each scenario a case may be made for the
ability of these seeds to result in the SMBHs we observe as
quasars, sometimes requiring the invocation of self-interacting
dark matter \citep{Ostriker2000,Huetal2006} or the accretion of
scalar fields \citep{BeanMagueijo2002}, there is as yet no
consensus on the matter.  We suggest that PBHs may be a viable
SMBH seed candidate because of their ability to build up large
dark matter halos that may
assist in further growth through baryon accretion later on.
Furthermore, forming earlier than Pop III stars, they have more
time to grow in the epoch before quasars are observed.

\subsection{Conclusions}
\label{sec:concl}

We have shown that primordial black holes can grow
significantly after formation through the accretion of
a dark matter halo.  In the radiation era
this can lead to an increase of total mass on order unity,
while during matter domination, the mass can grow by roughly
two orders of magnitude.  Although a dark matter halo
may not significantly increase the mass of the seed black
hole itself due to
the lack of a mechanism to dissipate angular momentum, the 
accumulation of a halo cannot be ignored when considering the
ability of a PBH to accrete gas in later eras (in the case of
collisional dark matter \citep{Ostriker2000}, the PBH can
directly accrete significant amounts of dark matter).  We have
also shown that the parameter space available to PBHs as
components of dark matter components is sufficient for them to
play an interesting role in galaxies.  PBHs may
be viable candidates for the seeds of intermediate mass black
holes, possibly responsible for ultra-luminous x-ray sources,
or they may play a role in seeding supermassive black holes
currently found in the centers of galaxies.

\section{Acknowledgements}

This material is based upon work supported under a National
Science Foundation Graduate Research Fellowship.  The authors
thank Niyaesh Afshordi, Bernard Carr, Andy Gould, Joshua Green,
Martin Rees, Ed Sirko, David Spergel, Paul Steinhardt, and everyone
at Underground Coffee for helpful feedback and suggestions.

\begin{appendix}

\section{Radiation era analytical approximation}
\label{sec:apprad}

The equation describing the dynamics of a
dark matter shell a distance $r$ from a black hole of mass $m_{BH}$
at time $t$ is:
\be
\frac{\ud^2 r}{\ud t^2} = -\frac{G m_{BH}}{r^2}-\frac{1}{4}\frac{r}{t^2}.
\ee
In the radiation era, we have $H = 1/(2t)$.  We define $r_i = r(t=t_i)$,
which leads to
\be
\frac{\ud r}{\ud t} (t=t_i) = H_i r_i = \frac{1}{2}\frac{r_i}{t_i}
\ee
where $H_i$ is the initial Hubble parameter.

We now consider the unperturbed solution, where $m_{BH}=0$.  This reduces
the above second-order equation to 
\be
\frac{\ud^2 r}{\ud t^2} = -\frac{1}{4}\frac{r}{t^2}.
\ee
We take the unperturbed behavior to be a powerlaw,
\be
r_0(t) = r_i\left(\frac{t}{t_i}\right)^\alpha.
\ee
Then,
\be
\frac{\ud^2 r_0}{\ud t^2} = \frac{r_i}{t_i^\alpha} \alpha (\alpha-1)t^{\alpha-2} = -\frac{1}{4}\frac{r_0}{t^2},
\ee
which we can solve to get
\be
\alpha = \frac{1}{2}.
\ee
This gives us the time evolution of the unperturbed solution:
\be
r_0 = \frac{r_i}{t_i^{1/2}}.
\ee
We can now take $r=r_0 + \delta r$.  Differentiating twice, we get
\be
\ddot{r} + \delta \ddot{r} = -\frac{G m_{BH}}{r_0^2\left(1+\frac{\delta r}{r_0}\right)^2} - \frac{1}{4} \frac{r_0}{t^2} - \frac{1}{4} \frac{\delta r}{t^2}.
\ee
Substituting in our solutions for $r_0$ and $\ddot{r_0}$, we obtain
\be
\delta \ddot{r} = -\frac{G m_{BH}}{r_0^2\left(1+\frac{\delta r}{r_0}\right)^2} - \frac{1}{4} \frac{\delta r}{t^2}.
\ee

The above equation is still exact, but we can take $\delta r/r$ to
be small to get the lowest order solution,
\be
\delta \ddot{r} + \frac{1}{4} \frac{\delta r}{t^2} = -\frac{G m_{BH}}{r_i^2} \frac{t_i}{t}.
\ee
For simplicity, we now define $x \equiv \delta r/r_i$ and
$\tau\equiv t/t_i$.  This gives us
\be
\ddot{x} + \frac{1}{4} \frac{x}{\tau^2} = -\frac{G m_{BH} t_i^2}{r_i^3 \tau}.
\ee
Defining
\be
\epsilon = \frac{G m_{BH} t_i^2}{r_i^3} \approx \left( \frac{\delta \rho}{\rho} \right)_i,
\ee
we have
\be
\ddot{x} + \frac{1}{4} \frac{x}{\tau^2} = -\frac{\epsilon}{\tau}.
\ee

The solution of the homogeneous equation, $\ddot{x} +x/(4\tau^2) = 0$,
is $x \propto \tau^\alpha$.  Finding the particular integral yields $x
= -4\epsilon\tau$, which gives a general solution
\be
x = A \tau^{1/2} - 4 \epsilon \tau
\ee
We solve for $A$ by considering that at $\tau=1$, $x=0$, which makes
$A = 4\epsilon$.  Thus we have
\be
x(\tau) = 4\epsilon (\tau^{1/2} - \tau) = \frac{\delta r}{r_i}(\tau).
\ee

For those shells of matter bound to the central black hole, each shell
will have a turnaround time (time at which the shell ceases to expand
in the Hubble flow and begins to fall back) and a collapse time (time
when the radius of the shell goes to zero).  

We define a shell's collapse time as the time when $r=0$:
\be
r = r_0 + \delta r = r_0 +\frac{\delta r}{r_i} r_i = 0.
\ee
Rewriting this with $x$ and $\tau$, we have
\be
r_i \tau_{coll}^{1/2} + x_{coll} r_i = 0,
\ee
and with the solution for $x$, this gives
\be
\tau_{coll} = \left( \frac{1+4\epsilon}{4\epsilon} \right)^2,
\ee
or
\be
t_{coll} = t_i \left( \frac{1+4\epsilon}{4\epsilon} \right)^2.
\ee

To find the amount of matter accreted by the black hole in the
radiation era, we choose some initial time, and find the amount
of matter accreted between that time and matter-radiation
equality.  This will likely be an overestimate, however, as any
interactions or other effects are more likely to slow accretion by
pulling matter away from the black hole.  This calculation will find
the amount of matter that had sufficient time to accrete, assuming
the accretion is steady and undisturbed.  Here we perform a rough
estimate of the amount of matter accreted.

We choose as the initial time when $z_i \approx 10^7$.
The final time is the time of matter-radiation equality, $t_f = t_{eq}$,
corresponding to $z_{eq} \approx 3 \times 10^3$.  During the radiation epoch,
$t \propto (1+z)^{-2}$, so we have $t_i/t_f \approx 10^{-7}$.  Setting $t_f =
t_{coll}$, we have
\be
\frac{t_{coll}}{t_i} = \left( \frac{1+4\epsilon_{min}}{4\epsilon_{min}} \right)^2 = 10^{7},
\ee
where $\epsilon_{min}$ reflects the fact that this is a maximal
estimate of the accretion.  Since we defined $\epsilon \equiv 
Gm_{BH}/(r_i^3 t_i^{-2})$, we can now write
\be
r^3_{i,max} = \frac{G m_{BH}}{t_i^{-2}\epsilon_{min}}.
\ee
Then,
\be
m_{acc,max} = \frac{4}{3} \pi r^3_{i,max} \rho_{m,i} = \frac{4}{3} \pi \frac{G m_{BH}}{t_i^{-2} \epsilon_{min}} \rho_{m,0} (1+z_i)^3,
\ee
which gives us
\be
\frac{m_{acc,max}}{m_{BH}} = \frac{4 \pi G \rho_{m,0}}{3 t_i^{-2} \epsilon_{min}} (1+z_i)^3.
\ee 
With some manipulation, this becomes
\be
\frac{m_{acc,max}}{m_{BH}} = \frac{2}{9} \left(\frac{t_i}{t_f}\right)^2 \left(\frac{t_f}{t_0}\right)^2 \frac{\Omega_{m,0}}{\epsilon_{min}}(1+z_i)^3
\ee
where the subscript 0 refers to the present era.  This is not exact, since
we are implicitly assuming that the present era is completely
matter-dominated, i.e.,
$\Omega_{m,0}=1$.  However, this can be neglected if we instead use
$z \approx 30$ as the final time -- in this case, the fraction $t_i/t_0$ is
not significantly changed and the factor $\Omega_{m,0}$ becomes unity.  For
$\epsilon_{min}$, we use $\delta \rho/\rho \approx 2.5 \times 10^{-5}$.
Here $t_f/t_0 = 6 \times 10^{-6}$, $t_i/t_f = 9 \times 10^{-8}$, and
$z_i = 10^7$,
which gives us
\be
\frac{m_{acc,max}}{m_{BH}} \approx 2.6.
\ee

Thus we see that the fractional increase in mass of the black hole in the
radiation era is on order 1.

\section{Matter era calculation with cosmological constant}
\label{sec:appmatt}

To calculate radial infall
of dark matter with the effect of the cosmological
constant included, we solve \citep{Lahav}:

\be
\frac{\ud^2 r}{\ud t^2} = -\frac{G M_i}{r^2} +
\frac{\Lambda r}{3},
\ee

The mass internal to a dark matter shell initially at $r_i$ is
given by
\begin{equation} \label{eq:mi}
M_i =
\frac{4}{3}\pi\rho_{m,i}r_i^3 + m_{BH},
\end{equation}
where $r_i$ is the initial physical radius, with
the initial density contrast defined as
\begin{equation} \label{eq:delta}
\Delta_{ci} = \frac{M_i}{(4/3)\pi r_i^3
\rho_{ci}} - 1.
\end{equation}
From this, \citet{Lahav} derive the
equation of motion:
\be \label{eq:dyn}
\frac{\ud^2 A}{\ud \tau^2} = -\frac{1}{2} (\Delta_{ci}+1)
A^{-2} + \lambda_i A,
\ee
where $\lambda_i = \Omega_{\Lambda,i}$, $\tau =
H_{i}t$, and where $A$ is the scale factor of the
shell,
$R(t) = A(t,r_i)r_i$. The initial conditions for the
Hubble flow are $A_i = 1$,
$\ud A/ \ud \tau = 1$.  This must be solved
numerically for the initial density contrast
$\Delta_{ci}$ corresponding to the shell collapsing
at the final time $\tau_f$. The mass accreted from
$\tau_i$ to $\tau_f$ is given by
\be \label{macc}
m_{acc} = m_{BH} \left(
\frac{1-\Omega_{\Lambda,i}-\Omega_{BH,i}}{\Delta_{ci}
+ \Omega_{\Lambda,i} + \Omega_{BH,i}} \right).
\ee

As a result of the presence of the cosmological
constant, there will be a last bound shell, which is
the last shell of matter that can in principle be
accreted by the black hole.  All shells internal to
this are bound and will turn around and fall back,
but those beyond it will continue expanding in the
Hubble flow.  Based on the calculation in \citet{SCO},
we find the last bound shell to be at an initial
radius of
\be
r_{\lambda}^3 = \frac{1}{2\pi}
\frac{m_{BH}}{\rho_{m,i}} \left(2\frac{\Omega_{\Lambda,i}}{\Omega_{m,i}}\right)^{-1/3},
\ee
where $\Omega_{m,i}$ is the initial density parameter in
matter.

The mass within the last bound shell is then
\be
m_\lambda  =  (4\pi/3)\rho_{m,i} r_\lambda^3,
\ee
and
\be
m_\lambda  =  \frac{2}{3} \left(
\frac{1-\Omega_{\Lambda,i}}{2 \Omega_{\Lambda,i}}
\right)^{1/3}  m_{BH}.
\ee

For the case of accretion beginning at matter-radiation
equality, the ratio of the mass within the last bound
shell to the initial mass of the PBH will be
$m_{\lambda}/m_{BH} \approx 1500$. This is the
mass increase possible, in principle, for each PBH.
However, as we derive this assuming each PBH to be
isolated, it may occur that the PBH will run out of
matter to accrete before this limit is reached, as it
has all been accreted by neighboring black holes or that
infall of much of this mass would occur after the present
epoch.  More realistically, as the PBHs grow, they will
begin to interact with one another, and their cluster
dynamics and mergers will have to be considered.  The
last bound shell mass can therefore be considered a
strict upper limit on the accretion.

\end{appendix}

\vfill 

\begin{thebibliography}{}

\bibitem[Afonso et al.(2003)]{Afonso2003}
  Afonso, C. et al. 2003, A\&A, 400, 951

\bibitem[Afshordi, McDonald \& Spergel(2003)]{AMS2003}
  Afshordi, N., McDonald, P., \& Spergel, D.N. 2003, ApJ,
594, 71L

\bibitem[Alcock et al.(2000)]{Alcock2000}
  Alcock, C. et al. 2000, ApJ, 542, 257

\bibitem[Alcock et al.(2001)]{Alcock2001}
  Alcock, C. et al. 2001, ApJ, 550, 169

\bibitem[Barth et al.(2003)]{Barthetal2003}
  Barth, A.J. et al. 2003, ApJ, 594, L95

\bibitem[Bean \& Magueijo(2002)]{BeanMagueijo2002}
  Bean, R. \& Magueijo, J. 2002, PhysRevD, 66, 063505

\bibitem[Berezin, Kuzmin \& Tkachev(1983)]{Berezin1983}
  Berezin, V.A., Kuzmin, V.A. \& Tkachev, I.I. 1983, PhysLettB, 120, 91

\bibitem[Bernardi et al.(2003)]{Bernardi2003}
  Bernardi, M. 2003, AJ, 125, 1849

\bibitem[Bertschinger(1985)]{Bert}
  Bertschinger, E. 1985, ApJ, 58, 39

\bibitem[Carr(2003)]{Carr2003}
  Carr, B.J. 2003, astro-ph/0310838

\bibitem[Carr(2005)]{Carr2005}
  Carr, B.J. 2003, astro-ph/0504034

\bibitem[Caldwell \& Casper(1996)]{CaldCasp1996}
  Caldwell, R. \& Casper, P. 1996, PhysRevD, 53, 3002

\bibitem[Carr \& Hawking(1974)]{CarrHawking}
  Carr., B.J. \& Hawking, S.W. 1974, MNRAS, 168, 399

\bibitem[Clifford(2004)]{Clifford2004}
  Clifford, W.M. 2004, ApJ, 611, 1080

\bibitem[Chisholm(2005)]{Chisholm2005}
  Chisholm, J.R. 2005, astro-ph/0509141

\bibitem[Crawford \& Schramm(1982)]{CrawfordSchramm1982}
  Crawford, M. \& Schramm, D.N. 1982, Nature, 298, 538

\bibitem[Cust\'{o}dio \& Horvath(1998)]{CustodioHorvath}
  Cust\'{o}dio, P.S. \& Horvath, J.E. 1998, PhysRevD, 58, 023504

\bibitem[Dalcanton et al.(1994)]{Dalcantonetal1994}
  Dalcanton, J.J. et al. 1994, ApJ, 424, 550

\bibitem[Dewangan, Titarchuk \& Griffiths(2005)]{Dewangan2005}
  Dewangan, G.C., Titarchuk, L. \& Griffiths, R.E. 2005, astro-ph/0509646

\bibitem[Dokuchaev, Eroshenko \& Rubin(2004)]{Dokuchaev2004}
  Dokuchaev, V.I., Eroshenko, Yu.N. \& Rubin, S.G., astro-ph/0412479

\bibitem[Fan et al.(2003)]{Fan2003}
  Fan, X. et al. 2003, AJ, 125, 1649

\bibitem[Fillmore \& Goldreich(1984)]{FillmoreGoldreich}
  Fillmore, J.A. \& Goldreich, P. 1984, ApJ, 281, 1

\bibitem[Fregeau et al.(2006)]{Fregeau2006}
  Fregeau, J.M. 2006, ApJ, 646, 135

\bibitem[Fryer \& Kalogera(2001)]{FryerKalogera2001}
  Fryer, C.L. \& Kalogera, V. 2001, ApJ, 554, 548

\bibitem[Garriga \& Sakellariadou(1993)]{GarSak1993}
  Garriga, J. \& Sakellariadou, M. 1993, PhysRevD, 48, 2502

\bibitem[Gould(2005)]{Gould2005}
  Gould, A., 2005, ApJ, 630, 891

\bibitem[Gunn \& Gott(1972)]{GG}
  Gunn, J.E. \& Gott, J.R. 1972, ApJ, 176, 1

\bibitem[Hawking(1971)]{Hawking1971}
  Hawking, S.W. 1971, MNRAS, 152, 75

\bibitem[Hawking(1975)]{Hawking1975}
  Hawking, S.W. 1975, Commun.Math.Phys., 43, 199

\bibitem[Hawking(1989)]{Hawking1989}
  Hawking, S.W. 1989, PhysLettB, 231, 273

\bibitem[Hawking, Moss \& Stewart(1982)]{HawkingMS1982}
  Hawking, S.W., Moss, I. \& Stewart, J. 1982, PhysRevD, 26, 2681

\bibitem[Hu et al.(2006)]{Huetal2006}
  Hu, J. et al., 2006, MNRAS, 365, 345

\bibitem[Ipser \& Sikivie(1984)]{IpserSikivie1984}
  Ipser, J. \& Sikivie, P. 1984, PhysRevD, 30, 712

\bibitem[Jedamzik(1997)]{Jedamzik1997}
  Jedamzik, K. 1997, PhysRevD, 55, 5871

\bibitem[Khlopov \& Polnarev(1980)]{KhlopovPolnarev1980}
  Khlopov, M.Yu. \& Polnarev, A.G. 1980, PhysLettB, 97, 383

\bibitem[Kohri \& Yokoyama(2000)]{Kohri2000}
  Kohri, K. \& Yokoyama, J. 2000, PhysRevD, 61, 023501

\bibitem[La \& Steinhardt(1989)]{LaStein1989}
  La, D. \& Steinhardt, P.J. 1989, PhysRevLett, 220, 375

\bibitem[Lacey \& Ostriker(1985)]{LaceyOstriker}
  Lacey, C.G. \& Ostriker, J.P. 1985, ApJ, 299, 633

\bibitem[Lahav et al.(1991)]{Lahav}
  Lahav, O. et al. 1991, MNRAS, 251, 128

\bibitem[Li et al.(2006)]{Lietal2006}
  Li, Y. et al. 2006, astro-ph/0608190

\bibitem[Lodato \& Natarajan(2006)]{LodatoNatarajan2006}
  Lodato, G. \& Natarajan, P. 2006, astro-ph/0606159

\bibitem[MacGibbon, Brandenberger \& Wichoski(1998)]{MacBrandWich1998}
  MacGibbon, J.H., Brandenberger, R.H. \& Wichoski, U.F. 1998,
  PhysRevD, 57, 2158

\bibitem[Madau \& Silk(2005)]{MadauSilk2005}
  Madau, P. \& Silk, J. 2005, MNRAS, 359, L37

\bibitem[Madhusudhan et al.(2006)]{Madhusudhanetal2006}
  Madhusudhan, N. et al. 2006, ApJ, 640, 918

\bibitem[Mapelli, Ferrara \& Rea(2006)]{Mapellietal2006}
  Mapelli, M., Ferrara, A. \& Rea, N. 2006, MNRAS, 368, 1340

\bibitem[Micic, Holley-Bockelmann \& Sigurdsson(2006)]{Micic2}
  Micic, M., Holley-Bockelmann, K. \& Sigurdsson, S. 2006,
astro-ph/0608493

\bibitem[Micic, Abel \& Sigurdsson(2005)]{MicicAbelSigurdsson2005}
  Micic, M., Abel, T. \& Sigurdsson, S. 2005, astro-ph/0512123

\bibitem[Mii \& Totani(2005)]{MiiTotani2005}
  Mii, H. \& Totani, T. 2005, ApJ, 628, 873

\bibitem[Miller(2004)]{Miller2004}
  Miller, J.M. 2004, astro-ph/0412526

\bibitem[Miller \& Hamilton(2002)]{Miller2002}
  Miller, M.C. \& Hamilton, D.P. 2002, MNRAS, 330, 232

\bibitem[Mushotzky(2004)]{Mushotzky2004}
  Mushotzky, R. 2004, Progress of Theoretical Physics
Supplement, 115, 27

\bibitem[Navarro, Frenk \& White(1997)]{NFW}
  Navarro, J.F., Frenk, C.S. \& White, S.D.M. 1997, ApJ, 490, 493

\bibitem[Nemiroff et al.(2001)]{Nemiroffetal2001}
  Nemiroff, R.J. et al. 2001, PhysRevLett, 86, 580

\bibitem[Niemeyer \& Jedamzik(1999)]{NiemeyerJedamzik}
  Niemeyer, J.C. \& Jedamzik, K. 1999, PhysRevD, 59, 124013

\bibitem[Novikov et al.(1979)]{Novikov1979}
  Novikov, I.D. et al. 1979, A\&A, 80, 104

\bibitem[Ostriker(2000)]{Ostriker2000}
  Ostriker, J.P. 2000, PhysRevLett, 84, 5258

\bibitem[Patruno et al.(2006)]{Patrunoetal2006}
  Patruno, A. et al. 2006, MNRAS, 370, L6

\bibitem[Polnarev \& Zemboricz(1988)]{PolnarevZemboricz1988}
  Polnarev, A.G. \& Zemboricz, R. 1988, PhysRevD, 43, 1106

\bibitem[Pooley \& Rappaport(2005)]{PooleyRappaport2005}
  Pooley, D. \& Rappaport, S. 2005, ApJ, 634, L85

\bibitem[Rubin, Khlopov \& Sakharov(2000)]{Rubin2000}
  Rubin, S.G., Khlopov, Yu.M. \& Sakharov, A.S. 2000,
Gravitation \& Cosmology Supplement, 6, 1

\bibitem[Ricotti, Ostriker \& Mack(2006)]{ROM}
  Ricotti, M., Ostriker, J.P. \& Mack, K.J. 2006, in
preparation.

\bibitem[Shapiro(2005)]{Shapiro2005}
  Shapiro, S.L. 2005, ApJ, 620, 59

\bibitem[So{\l}tan(1982)]{Soltan}
  So{\l}ltan, A. 1982, MNRAS, 200, 115

\bibitem[Spaans \& Silk(2006)]{SpaansSilk2006}
  Spaans, M. \& Silk, J. 2006, astro-ph/0601714

\bibitem[Spergel et al.(2006)]{Spergel2006}
  Spergel, D.N. et al. 2006, astro-ph/0603449

\bibitem[Subramanian, Cen \& Ostriker(2000)]{SCO}
  Subramanian, K., Cen, R. \& Ostriker, J.P. 2000, ApJ,
538, 528

\bibitem[Volonteri \& Rees(2005)]{VR}
  Volonteri, M. \& Rees, M.J. 2005, astro-ph/0506040

\bibitem[Wambsganss(2002)]{Wambsganss2002}
  Wambsganss, J. 2002, astro-ph/0207616

\bibitem[Wilkinson et al.(2001)]{Wilkinsonetal2001}
  Wilkinson, P.N. et al. 2001, PhysRevLett, 86, 584

\bibitem[Willott, McLure \& Jarvis(2003)]{Willottetal2003}
  Willott, C.J., McLure, R.J. \& Jarvis, M.J. 2003, ApJ, 587, L15

\bibitem[Winter, Mushotzky \& Reynolds(2006)]{Winteretal2006}
  Winter, L.M., Mushotzky, R. \& Reynolds, C.S. 2006, astro-ph/0512480

\bibitem[Yoo, Chanam\'{e} \& Gould(2004)]{YooChanameGould2004}
  Yoo, J., Chanam\'{e}, J. \& Gould, A. 2004, ApJ, 601, 311

\bibitem[Yu \& Tremaine(2002)]{YuTremaine}
  Yu, Q \& Tremaine, S. 2002, MNRAS, 335, 965

\bibitem[Zel'dovich \& Novikov(1967)]{ZN1967}
  Zel'dovich, Ya.B. \& Novikov, I.D. 1967, Sov. Astron. A. J.,
10, 602


\end{thebibliography}
\end{document}